\documentclass[12pt]{article}

\usepackage[dvips]{graphicx}

\def\dspace{\baselineskip = 0.30in}

\def\lapproxeq{\lower .7ex\hbox{$\;\stackrel{\textstyle
<}{\sim}\;$}}
\def\gapproxeq{\lower .7ex\hbox{$\;\stackrel{\textstyle
>}{\sim}\;$}}

\begin{document}

\dspace

\begin{titlepage}
\begin{flushright}
BA-02-40\\
\end{flushright}
\vskip 2cm
\begin{center}
{\Large\bf
Inflationary Cosmology with \\
Five Dimensional $SO(10)$  
}
\vskip 1cm
{\normalsize\bf
Bumseok Kyae\footnote{bkyae@bartol.udel.edu} and
Qaisar Shafi\footnote{shafi@bxclu.bartol.udel.edu}
}
\vskip 0.5cm
{\it Bartol Research Institute, University of Delaware, \\Newark,
DE~~19716,~~USA\\[0.1truecm]}

%

\end{center}
\vskip .5cm


\begin{abstract}

We discuss inflationary cosmology in   
a five dimensional $SO(10)$ model compactified 
on $S^1/(Z_2\times Z_2')$, which yields 
$SU(3)_c\times SU(2)_L\times U(1)_Y\times U(1)_X$ 
below the compactification scale.   
The gauge symmetry $SU(5)\times U(1)_X$ is preserved on one 
of the fixed points, while ``flipped'' $SU(5)'\times U(1)'_X$ is 
on the other fixed point.   
Inflation is associated with $U(1)_X$ breaking, 
and is implemented through $F$-term scalar potentials 
on the two fixed points.    
A brane-localized Einstein-Hilbert term allows both branes  
to have positive tensions during inflation.  
The scale of $U(1)_X$ breaking is fixed from $\delta T/T$ measurements 
to be around $10^{16}$ GeV, and 
the scalar spectral index $n=0.98-0.99$.  
The inflaton field decays into right-handed neutrinos whose subsequent 
out of equilibrium decay yield the observed baryon asymmetry 
via leptogenesis.

\end{abstract}
\end{titlepage}

\newpage

\section{Introduction}

In a recent paper~\cite{KS},
we showed how supersymmetric (SUSY) inflation can be realized
in five dimensional (5D) models in which the fifth dimension is compactified on
the orbifold $S^1/(Z_2\times Z_2')$.
Orbifold symmetry breaking in higher dimensional grand unified theories (GUTs) 
have recently attracted a great deal of attention   
because 
of the apparent ease with which they can circumvent 
two particularly pressing problems encountered 
in four dimensional (4D) SUSY GUTs~\cite{orbifold}:  
namely, the doublet-triplet (DT) splitting problem 
and the problem caused by dimension five nucleon decay. 
The apparent reluctance of the proton decay, 
as shown by the recent lower limits on its lifetime~\cite{proton},     
seems to be in broad disagreement with the predicted rates 
from dimension five processes in minimal SUSY $SU(5)$ and $SO(10)$ models. 
The existence of the orbifold dimension makes it possible to implement 
DT splitting and simultaneously suppress (or even eliminate) 
dimension five proton decay.  

The inflationary scenario described in Ref.~\cite{KS} was inspired 
by the above particle physics considerations and 
has some novel features.  
The primordial density (temperature) fluctuations are proportional to
$(M/M_{\rm Planck})^2$, along the lines of the 4D model 
proposed in Ref.~\cite{hybrid}.   
Here $M$ refers to the scale of some symmetry breaking that is
associated with inflation, and $M_{\rm Planck}=1.2\times 10^{19}$ GeV 
denotes the Planck scale.   
In an $SO(10)$ model, for example, the orbifold breaking can be used to yield 
the subgroup $H=SU(4)_c\times SU(2)_L\times SU(2)_R$~\cite{dermisek}, 
so that inflation is associated with the breaking of $H$ 
to the minimal supersymmetric standard model (MSSM) 
gauge group~\cite{khalil,KS}.  
The anisotropy measurements~\cite{cobe} can provide a determination of $M$ 
independently of any particle physics considerations. 
$M$ turns out to be quite close (or equal) to the SUSY GUT scale 
of around $10^{16}$ GeV.  
Last but not least, the scalar spectral index of density fluctuations is
very close to unity ($n=0.98-0.99$)~\cite{khalil,KS}.
The gravitational wave contribution to the quadrupole anisotropy is found
to be essentially negligible.  
In an $SO(10)$ model with the subgroup $H$ given above, 
the inflaton decays into the MSSM singlet (right-handed) neutrinos,
whose out of equilibrium decay leads to the observed baryon asymmetry via
leptogenesis~\cite{lepto,ls}.   
As we will see, 
this is also the case even with a different subgroup $H$ of $SO(10)$.
Baryogenesis via leptogenesis appears to be a rather generic feature
of 5D $SO(10)$ based inflationary models considered here.  

Although our considerations are quite general, in this paper we focus on
an example based on $SO(10)$ with subgroup 
$SU(3)_c\times SU(2)_L\times U(1)_Y\times U(1)_X$ obtained from  
$S^1/(Z_2\times Z_2')$ orbifolding.  
The standard $SU(5)\times U(1)_X$ is preserved on one brane, while
``flipped'' $SU(5)'\times U(1)'_X$ is on the other brane.  
All massless modes from the chiral component of the 5D vector multiplet
can be easily superheavy 
by introducing Higgs fields in the bulk.      
Inflation is associated with the breaking of $U(1)_X$, followed by its decay 
into right-handed neutrinos, which subsequently generate 
a primordial lepton asymmetry.
The gravitino constraint on the reheat temperature~\cite{gravitino} 
imposes important constraints on the masses of the right-handed neutrinos
which can be folded together with the information now available
from the oscillation experiments~\cite{nuoscil}.

As emphasized in Ref.~\cite{KS}, implementation in five dimensions of the 
inflationary scenario considered in Ref.~\cite{hybrid} requires some care.  
Note that the five dimensional setup is the appropriate one because of 
the proximity of the scale of inflation and the comapactification scale 
(both are of order $M_{GUT}$).  
The inflaton potential must be localized on the orbifold fixed points 
(branes), since a superpotential in the bulk is not allowed.  
For a vanishing bulk cosmological constant, a three space inflationary 
solution triggered by non-zero brane tensions (or vacuum energies) 
exists~\cite{KS}. 
However, 5D Einstein equations often require that the signs of the brane 
tensions on the two branes are opposite, which  is undesirable.  
As shown in Ref.~\cite{KS}, this problem can be circumvented 
by introducing a brane-localized
Einstein-Hilbert term in the action.
The two brane tensions are both positive during inflation, 
and they vanish when it ends.
%
%

The plan of our paper is as follows. In section 2, we review cosmology 
in a five dimensional setting, and discuss in particular the transition from 
an inflationary to the radiation dominated epoch.  
In section 3, we discuss the orbifold breaking of $SO(10)$ 
to $SU(3)_c\times SU(2)_L\times U(1)_Y\times U(1)_X$.  
Section 4 summarizes the salient features of the inflationary scenario and 
subsequent leptogenesis.  
Our conclusions are summarized in section 5.

\section{5D Cosmology}


We consider 5D space-time $x^M=(x^{\mu},y)$, $\mu=0,1,2,3$, 
compactified on an $S^1/Z_2$ orbifold, and 
the supergravity (SUGRA) action is typically given by
\begin{eqnarray} \label{action}
S=\int d^4x \int_{-y_c}^{y_c}dy~e\bigg[\frac{M_5^3}{2}~R_5+{\cal L}_B
+\sum_{i=I,II}\frac{\delta(y-y_i)}{e_5^5}\bigg(\frac{M_i^2}{2}~\bar{R}_4
+{\cal L}_i\bigg)
\bigg] ~, 
\end{eqnarray}
where $R_5$ ($\bar{R}_4$) stands for the 5 dimensional (4 dimensional) 
Einstein-Hilbert term, ${\cal L}_B$ (${\cal L}_I$, ${\cal L}_{II}$)
denotes some unspecified bulk (brane) 
contributions to the full Lagrangian, and $y_I=0$, $y_{II}=y_c$ indicate  
the brane positions.  
The brane scalar curvature term $\bar{R}_4(\bar{g}_{\mu\nu})$
is defined through the induced metric,
$\bar{g}_{\mu\nu}(x)\equiv g_{\mu\nu}(x,y=0)$ ($\mu,\nu=0,1,2,3$).
The brane-localized Einstein-Hilbert terms\footnote{
The importance of the
brane-localized 4D Einstein-Hilbert term, especially for generating 4D gravity
in a higher dimensional non-compact flat space was first noted
in Ref.~\cite{braneR}. 
} in Eq.~(\ref{action})
are allowed also in SUGRA, but should, of course, be accompanied 
by brane gravitino kinetic terms as well as other terms, 
as spelled out in the off-shell SUGRA formalism \cite{kyae}.
Here we assume that the bulk cosmological constant is zero.

For the cosmological solution let us take the following metric ansatz,
\begin{eqnarray} \label{metric}
ds^2=\beta^2(t,y)\bigg(-dt^2+a^2(t)~d\vec{x}^2\bigg)+dy^2 ~,
\end{eqnarray}
which shows that the three dimensional space is homogeneous and isotropic.  
The non-vanishing components of the 5D Einstein tensor derived 
from Eq.~(\ref{action}) are 
\begin{eqnarray}
G^0_0&=&3\bigg[\bigg(\frac{\beta''}{\beta}\bigg)
+\bigg(\frac{\beta'}{\beta}\bigg)^2~\bigg]
-\frac{3}{\beta^2}\bigg[\bigg(\frac{\dot{\beta}}{\beta}
+\frac{\dot{a}}{a}\bigg)^2~\bigg] \nonumber \\ 
&&-\sum_{i=I,II}\delta(y-y_i)\frac{M_i^2}{M_5^3}\frac{3}{\beta^2}\bigg[
\bigg(\frac{\dot{\beta}}{\beta}
+\frac{\dot{a}}{a}\bigg)^2~\bigg]
~, \label{eins00} \\
G^i_i&=&3\bigg[\bigg(\frac{\beta''}{\beta}\bigg)
+\bigg(\frac{\beta'}{\beta}\bigg)^2~\bigg]
-\frac{1}{\beta^2}\bigg[2\frac{\ddot{\beta}}{\beta}+2\frac{\ddot{a}}{a}
+4\frac{\dot{\beta}}{\beta}\frac{\dot{a}}{a}
-\bigg(\frac{\dot{\beta}}{\beta}\bigg)^2+\bigg(\frac{\dot{a}}{a}\bigg)^2
~\bigg] \nonumber \\
&&-\sum_{i=I,II}\delta(y-y_i)\frac{M_i^2}{M_5^3}\frac{1}{\beta^2}\bigg[
2\frac{\ddot{\beta}}{\beta}+2\frac{\ddot{a}}{a}
+4\frac{\dot{\beta}}{\beta}\frac{\dot{a}}{a}
-\bigg(\frac{\dot{\beta}}{\beta}\bigg)^2+\bigg(\frac{\dot{a}}{a}\bigg)^2
~\bigg]
~, \label{einsii} \\
G^5_5&=&6\bigg[\bigg(\frac{\beta'}{\beta}\bigg)^2~\bigg]-\frac{3}{\beta^2}
\bigg[\frac{\ddot{\beta}}{\beta}+\frac{\ddot{a}}{a}
+3\frac{\dot{\beta}}{\beta}\frac{\dot{a}}{a}+\bigg(\frac{\dot{a}}{a}\bigg)^2
~\bigg]
~, \label{eins55} \\
G_{05}&=&-3\bigg[\bigg(\frac{\beta'}{\beta}\bigg)^{\cdot}~ \bigg] ~,
\label{eins05}
\end{eqnarray}
where primes and dots respectively denote derivatives with respect to 
$y$ and $t$, and the terms accompanied by delta functions 
arise from the brane-localized Einstein-Hilbert terms.

Let us first discuss inflation under this setup.
Since 5D $N=1$ SUSY does not allow a superpotential
(and the corresponding $F$-term scalar potential) in the bulk, 
we introduce the inflaton scalar potentials $V_{I,II}(\phi)$ ($\geq 0$) 
on the two branes where only 4D $N=1$ SUSY is preserved~\cite{KS,king}.   
The energy-momentum tensor during inflation is given by  
\begin{eqnarray}
T^0_0~=~T^i_i&=&-\delta(y)\frac{V_{I}}{M_5^3}
-\delta(y-y_c)\frac{V_{II}}{M_5^3} ~,   \\
T^5_5&=&0 ~,
\end{eqnarray}
where $i=1,2,3$, and $V_{I}$, $V_{II}$ are the scalar potentials 
on the branes at $y=0$ (B1) and $y=y_c$ (B2), respectively.   
They should be suitably chosen   
to provide a large enough number of e-foldings to resolve the horizon and 
flatness problems.  
The end of inflation is marked by the breaking of the ``slow roll'' conditions,
and the inflaton rolls quickly to the true suepersymmetric vacuum  
with flat 4D space-time.  
Thus, for the inflationary epoch 
it is sufficient to consider only scalar potentials 
in the energy-momentum tensor. 
We will discuss more general cases later.    
%
%

The exact inflationary solution is~\cite{KS}
\begin{eqnarray}
&&\beta(y)=H_0|y|+c ~, \label{beta} \\
&&a(t)=e^{H_0t} ~, \label{a}
\end{eqnarray}
where $H_0$ ($>0$) is the Hubble constant during inflation.  
%
%
The integration constant $c$ in Eq.~(\ref{beta}) can be normalized 
to unity without loss of generality.  
%
%
%
The introduction of the brane-localized Einstein-Hilbert terms 
do not affect the bulk solutions, Eqs.~(\ref{beta}) and (\ref{a}),
but they modify the boundary conditions.
The solution $\beta(y)$ should satisfy the following boundary conditions
at $y=0$ and $y=y_c$,
\begin{eqnarray} \label{bdy1}
\frac{V_{I}}{6M_5^3}&=&-H_0+\frac{1}{2}\frac{M_I^2}{M_5^3}~H_0^2~, \\
\frac{V_{II}}{6M_5^3}&=&\frac{H_0}{1+H_0y_c}
+\frac{1}{2}\frac{M_{II}^2}{M_5^3}~\frac{H_0^2}{(1+H_0y_c)^2}~.   
\label{bdy2}
\end{eqnarray}
Thus, $H_0$ and $y_c$ are determined by $V_I$ and $V_{II}$.
Note that the brane cosmological constants (scalar potentials) $V_I$
and $V_{II}$ are related to the Hubble constant $H_0$.
While the non-zero brane cosmological constants are responsible for
inflating the 3-space, their subsequent vanishing 
restores SUSY and guarantees the flat 4D space-time.  
Since $V_{II}$ must be zero when $V_I=0$,
it is natural that the scalar field controlling the end of inflation is
introduced in the bulk.  
With SUSY broken at low energies, the minima of the inflaton potentials
on both branes should be fine-tuned to zero~\cite{kkl2,selftun}.    

From Eqs.~(\ref{bdy1}) and (\ref{bdy2}) we note that 
in the absence of the brane-localized Einstein-Hilbert term at $y=0$, 
the inflaton potentials (brane cosmological constants) 
$V_I$ and $V_{II}$ should have opposite signs.
However, a suitably large value of $M_I/M_5$~\cite{braneR} 
can flip the sign of $V_I$~\cite{KS}, so that both $V_I$ and 
$V_{II}$ are positive.  
%
%
Thus, a brane-localized Einstein-Hilbert term at $y=0$ seems essential 
for successful $F$-term inflation in the 5D SUSY framework.
Its introduction does not conflict with any symmetry, 
and in Ref.~\cite{KS} a simple model for realizing a large ratio $M_I/M_5$ 
was proposed.  

The 4D reduced Planck mass ($\equiv (M_{\rm Planck}/8\pi)^{1/2}$) 
is given by  
%
\begin{eqnarray}
M_{P}^2&=& M_5^3\int_{-y_c}^{y_c}dy\beta^2+M_I^2+M_{II}^2  \nonumber \\
&=&M_5^3y_c\bigg(\frac{2}{3}H_0^2y_c^2+2H_0y_c+2\bigg)+M_I^2+M_{II}^2 ~,    
\end{eqnarray}
while the 4D effective cosmological constant is calculated to be 
\begin{eqnarray}
\Lambda_{\rm eff}&=&\int_{-y_c}^{y_c}dy\beta^4\bigg[M_5^3\bigg(
4\bigg(\frac{\beta''}{\beta}\bigg)+6\bigg(\frac{\beta'}{\beta}\bigg)^2\bigg)
+\delta(y)V_I+\delta(y-y_c)V_{II}\bigg]  \nonumber \\
&=&3H_0^2\bigg[M_5^3y_c\bigg(\frac{2}{3}H_0^2y_c^2
+2H_0y_c+2\bigg)+M_I^2+M_{II}^2\bigg] \nonumber \\
&=&3H_0^2M_P^2 ~,  
\end{eqnarray}
which vanishes when $V_I=V_{II}=0$.

%
%
After inflation, the inflaton decays into brane 
and (subsequently) bulk fields, 
which reheat the whole 5 dimensional universe. 
To quantify the inflaton and radiation (or matter) dominated epochs, 
we use the fluid approximation, 
\begin{eqnarray} \label{fluid} 
T^{M}\,_N=\frac{1}{M_5^3}
\left(\begin{array}{cccc|c}
-\rho & 0 & 0 & 0 & T^0\,_{5}\\
0 & p & 0 & 0 & 0 \\
0 & 0 & p & 0 & 0 \\
0 & 0 & 0 & p & 0 \\ \hline
T^5\,_{0} & 0 & 0 & 0 & P_5
\end{array}\right) ~,  
\end{eqnarray} 
%
%
where $\rho$ and $p$ are contributed by bulk and brane matter,  
\begin{eqnarray} \label{matter}
\rho&\equiv&\frac{1}{2y_c}\rho_B+\sum_{i=I,II}\delta(y-y_i)\rho_{i} ~,\\
p&\equiv&\frac{1}{2y_c}p_B+2y_c\sum_{i=I,II}\delta(y-y_i)p_i ~. 
\label{matter2}
\end{eqnarray}
Note that in Eq.~(\ref{fluid}) the non-zero off-diagonal components, 
$T^0\,_5$ ($=\frac{-1}{\beta^2}T_{05}$) and $T^5\,_0$ ($=T_{05}$) 
are considered.    
In Eqs.~(\ref{matter}) and (\ref{matter2}), 
we normalize $\rho_B$ and $p_B$ with the circumference of
the extra dimension, so their components have the same mass dimension 
as their brane counterparts.  
%
%
%
%
With Eqs.~(\ref{eins00})--(\ref{eins05}) and (\ref{fluid}),   
the 5D ``Friedmann-like'' equations are readily written, 
\begin{eqnarray}
\frac{1}{2y_cM_5^3}~\rho_B&=&\frac{3}{\beta^2}\bigg[
\bigg(\frac{\dot{\beta}}{\beta}\bigg)^2+2\frac{\dot{\beta}}{\beta}H
+\bigg(H^2-h^2\bigg)\bigg] ~,  \label{rhoB}  \\
\frac{1}{M_5^3}~\rho_{I}&=&\bigg[3\frac{M_I^2}{M_5^3}H^2-6h\bigg] ~, \\
\frac{1}{M_5^3}~\rho_{II}&=&\bigg[\frac{M_{II}^2}{M_5^3}
\frac{3}{\beta^2}\bigg(\bigg(\frac{\dot{\beta}}{\beta}\bigg)^2
+2\frac{\dot{\beta}}{\beta}H+H^2\bigg)+6\frac{h}{\beta}\bigg]_{y=y_c} ~,  \\
\frac{1}{2y_cM_5^3}~p_B&=&-\frac{1}{\beta^2}\bigg[
2\frac{\ddot{\beta}}{\beta}-\bigg(\frac{\dot{\beta}}{\beta}\bigg)^2
+4\frac{\dot{\beta}}{\beta}H+2\dot{H}+3\bigg(H^2-h^2\bigg)\bigg] ~, \\
\frac{1}{M_5^3}~p_{I}&=&-\bigg[\frac{M_I^2}{M_5^3}\bigg(
2\dot{H}+3H^2\bigg)-6h\bigg]  ~, \\
\frac{1}{M_5^3}~p_{II}&=&-\bigg[\frac{M_{II}^2}{M_5^3}\frac{1}{\beta^2}\bigg(
2\frac{\ddot{\beta}}{\beta}-\bigg(\frac{\dot{\beta}}{\beta}\bigg)^2
+4\frac{\dot{\beta}}{\beta}H
+2\dot{H}+3H^2\bigg)+6\frac{h}{\beta}\bigg]_{y=y_c} ~, \\
\frac{1}{2y_cM_5^3}~P_5&=&-\frac{3}{\beta^2}\bigg[
\frac{\ddot{\beta}}{\beta}+3\frac{\dot{\beta}}{\beta}H+\dot{H}
+2\bigg(H^2-h^2\bigg)\bigg] ~, \\
\frac{1}{2y_cM_5^3}~T_{05}&=&-3~{\rm sgn}(y)\bigg[\frac{\dot{h}}{\beta}
-\frac{h}{\beta}\frac{\dot{\beta}}{\beta}\bigg] ~. \label{T05}
\end{eqnarray}
Here ${\rm sgn}(y)\equiv 1(-1)$ for $y>0(<0)$, and  
\begin{eqnarray}
H(t)&\equiv& \frac{\dot{a}}{a} ~, \\
\beta(t,y)&=&h(t)|y|+1 ~.  
\end{eqnarray}
For $M_I>M_5, M_{II}$, and $H>>h$,  
the brane matter contribution from B1 is dominant, and 
Eqs.~(\ref{rhoB})--(\ref{T05}) reduce to 
the approximate 4D Friedmann equations, 
\begin{eqnarray} \label{fried1}
&&\bigg(\frac{\dot{a}}{a}\bigg)^2\approx\frac{1}{3M_4^2}~\rho_I~, \\
&&~~\frac{\ddot{a}}{a}\approx\frac{-1}{6M_4^2}\bigg(\rho_I+3p_I
\bigg)~. \label{fried2}
\end{eqnarray}
Eqs.~(\ref{rhoB})--(\ref{T05}) satisfy 
the energy-momentum conservation law, 
$\nabla_MT^M_N=0$ whose $N=0$ and $N=5$ components are~\cite{kim-kyae}    
\begin{eqnarray} \label{conserv1}
\dot{\rho}+3\bigg(\frac{\dot{\beta}}{\beta}+H\bigg)(\rho+p)&=&
T^{5'}\,_0+4\frac{\beta'}{\beta}T^{5}\,_0  
\nonumber \\
&=&2y_cM_5^3\bigg[G_{05}'+4\frac{\beta'}{\beta}G_{05}\bigg] ~, 
\\ \label{conserv2}
P_5'+\frac{\beta'}{\beta}\bigg(4P_5-3p+\rho\bigg)&=&
-\dot{T}^{0}\,_5-\bigg(4\frac{\dot{\beta}}{\beta}+3H\bigg)T^{0}\,_5  
\nonumber \\
&=&2y_cM_5^3\frac{1}{\beta^2}\bigg[\dot{G}_{05}+\bigg(
2\frac{\dot{\beta}}{\beta}+3H\bigg)G_{05}\bigg] ~. 
\end{eqnarray}
The inflaton contributes to the energy momentum tensor, Eq.~(\ref{fluid}),    
\begin{eqnarray} \label{EM}
T_{MN}\equiv ~T^{\rm inf}_{MN}+T^{\rm m}_{MN} ~,  
\end{eqnarray}
where $T^{\rm inf}_{MN}$ denotes the contributions to 
the energy momentum tensor from the inflaton $\phi(t,y)$, 
\begin{eqnarray} \label{inflaton}
T^{\rm inf}_{MN}&\equiv& \frac{1}{2y_c}\partial_M\phi\partial_N\phi
-\frac{1}{4y_c}g_{MN}\partial^P\phi\partial_P\phi  \\
&&+\sum_{i=I,II}\delta(y-y_i)\delta_M^\mu\delta_N^\nu\bigg[
\partial_{\mu}\phi\partial_\nu\phi
-g_{\mu\nu}\bigg(\frac{1}{2}\partial^\lambda\phi\partial_\lambda\phi 
+V_i(\phi)\bigg)\bigg]~,  \nonumber
\end{eqnarray}
and $T^{\rm m}_{MN}$ is assumed to have the same form as Eq.~(\ref{fluid}).  
%
%
The conservation law $\nabla^MT^{\rm inf}_{MN}=0$ 
gives rise to the scalar field equation   
in the presence of both the brane and bulk kinetic terms.   
If only the inflaton potentials on the branes, $V_I(\phi)$ and $V_{II}(\phi)$
are dominant in Eq.~(\ref{EM}),  one can check that 
the solutions reduce to Eqs.~(\ref{beta}) and (\ref{a}), namely,
$H=h={\rm constant}~(=H_0)$.
The inflaton decay produces $T^{\rm m}_{MN}$.

%
%

We have tacitly assumed that the interval separating the two branes
(orbifold fixed points) remains fixed during inflation.  
The dynamics of the orbifold fixed points, 
unlike the D-brane case~\cite{Dbrane},
is governed only by the $g_{55}(x,y)$ component of the metric tensor.
The real fields $e_5^5$, $B_5$, and the chiral fermion $\psi^2_{5R}$
in 5D gravity multiplet are assigned even parity under $Z_2$ \cite{kyae},
and they compose an $N=1$ chiral multiplet on the branes.
The associated superfield can acquire a superheavy mass and
its scalar component can develop a vacuum expectation value (VEV) 
on the brane.
With superheavy brane-localized mass terms,
the low-lying Kaluza-Klein (KK) mass spectrum is shifted 
so that even the lightest mode 
obtains a compactification scale mass~\cite{localmass}. 
Since this mass is much greater than $H_0$ 
the interval distance is stable even during inflation.  
This stabilization of the interval distance in turn leads to 
the stabilization of the warp factor $\beta(y)$.  
This is because the fluctuation $\delta \beta(y)$
of the warp factor near the solution
in Eq.~(\ref{beta}) turns out to be proportional to
the interval length variation $\delta g_{55}$
from the linearized 5D Einstein equation~\cite{chacko}.

So far we have discussed only $S^1/Z_2$ orbifold compactification.
The results can be directly extended to $S^1/(Z_2\times Z_2')$.  
Within the framework discussed in this section,
we can accommodate any promising 4D SUSY inflationary model.
We consider one particular model below which comes from compactifying 
$SO(10)$ on an $S^1/(Z_2\times Z_2')$.  

%
%
%

\section{${\bf SU(3)_c\times SU(2)_L\times U(1)_Y\times U(1)_X}$
Model}

%

We consider the $N=2$ (in the 4D sense) SUSY $SO(10)$ model in 5D space-time, 
where the 5th dimension is compactified on an $S^1/(Z_2\times Z_2')$.  
$Z_2$ reflects $y\rightarrow -y$, and $Z_2'$ reflects
$y'\rightarrow -y'$, with $y'=y+y_c/2$.
There are two independent orbifold fixed points (branes)
at $y=0$ and $y=y_c/2$.
The $S^1/(Z_2\times Z_2')$ orbifold compactification is exploited to yield 
$N=1$ SUSY as well as break $SO(10)$ to some suitable subgroup.   
 
Under $SU(5)\times U(1)_X$, the $SO(10)$ generators are split into  
\footnote{The $SO(2n)$ generators are represented as
$\left(\begin{array}{cc}
A+C&B+S\\
B-S&A-C
\end{array}\right)$, where $A$,$B$, $C$ are $n\times n$ anti-symmetric matrices
and $S$ is an $n\times n$ symmetric matrix \cite{zee}.  
By a unitary transformation,
the generators are given by
$\left(\begin{array}{cc}
A-iS&C+iB\\
C-iB&A+iS
\end{array}\right)$, where $A$ and $S$ denote $U(n)$ generators, and
$C\pm iB$ transform as $n(n-1)/2$ and $\overline{n(n-1)/2}$
under $SU(n)$.}
\begin{eqnarray} \label{so10}
T_{SO(10)}=\left[\begin{array}{c|c}
{\bf 24}_0+{\bf 1}_0& {\bf 10}_{-4}\\
\hline
{\bf \overline{10}}_{4}&{\bf \overline{24}}_0-{\bf 1}_0
\end{array}\right]_{10\times 10} ~, 
\end{eqnarray}
where the subscripts labeling the $SU(5)$ representations indicate   
$U(1)_X$ charges, and the subscript ``$10\times 10$'' denotes the matrix 
dimension.  Also, ${\bf 24}$ ($={\bf \overline{24}}$) corresponds 
to $SU(5)$ generators while
${\rm diag}~({\bf 1}_{5\times 5},-{\bf 1}_{5\times 5})$
is the $U(1)_X$ generator.
The $5\times 5$ matrices ${\bf 24}_0$ and ${\bf 10}_{-4}$ are further 
decomposed under $SU(3)_c\times SU(2)_L\times U(1)_Y$ as 
\begin{eqnarray} \label{24}
&&{\bf 24}_{0}=\left(\begin{array}{cc}
{\bf (8,1)}_{0}+{\bf (1,1)}_0 & {\bf (3,\overline{2})}_{-5/6}\\
{\bf (\overline{3},2)}_{5/6} & {\bf (1,3)}_{0}-{\bf (1,1)}_0
\end{array}\right)_{0}~,  \\
&&{\bf 10}_{-4}=\left(\begin{array}{cc}
{\bf (\overline{3},1)}_{-2/3} & {\bf (3,2)}_{1/6}\\
{\bf (3,2)}_{1/6} & {\bf (1,1)}_{1}
\end{array}\right)_{-4} ~,
\end{eqnarray}
%
%
Thus, each representation carries two independent $U(1)$ charges.
Note that the two ${\bf (3,2)}_{1/6}$s in ${\bf 10}_{-4}$ are identified.  
%
%

Consider the two independent $Z_2$ and $Z_2'$ group elements,
\begin{eqnarray} \label{p1}
P&\equiv&{\rm diag.}\bigg(I_{3\times 3},~I_{2\times 2},
-I_{3\times 3},-I_{2\times 2}\bigg)~, \\
P'&\equiv&{\rm diag.}\bigg(-I_{3\times 3},~I_{2\times 2},
~I_{3\times 3},-I_{2\times 2}\bigg)~,  \label{p2}
\end{eqnarray}
which satisfy $P^2=P'^2={\bf 1}_{5\times 5}$.
Under the operations $PT_{SO(10)}P^{-1}$ and $P'T_{SO(10)}P^{'-1}$,
the matrix elements of $T_{SO(10)}$ transform as 
\begin{eqnarray} \label{so10/z2z2}
\left[\begin{array}{cc|cc}
{\bf (8,1)}_{0}^{++} &
~{\bf (3,\overline{2})}_{-5/6}^{+-} &
{\bf (\overline{3},1)}_{-2/3}^{--} & {\bf (3,2)}_{1/6}^{-+} \\
{\bf (\overline{3},2)}_{5/6}^{+-} &
{\bf (1,3)}_{0}^{++} &
{\bf (3,2)}_{1/6}^{-+} & {\bf (1,1)}_{1}^{--} \\
\hline
~{\bf (3,1)}_{2/3}^{--} & ~{\bf (\overline{3},\overline{2})}_{-1/6}^{-+} &
{\bf (8,1)}_{0}^{++}
& {\bf (\overline{3},2)}_{5/6}^{+-} \\
~{\bf (\overline{3},\overline{2})}_{-1/6}^{-+} & ~{\bf (1,1)}_{-1}^{--} &
~{\bf (3,\overline{2})}_{-5/6}^{+-} &
{\bf (1,3)}_{0}^{++}
\end{array}\right]_{10\times 10} ~,
\end{eqnarray}
where the superscripts of the matrix elements indicate the eigenvalues of the
$P$ and $P'$ operations.
Here we omitted the two $U(1)$ generators (${\bf (1,1)_0^{++}}$s) 
to avoid too much clutter.  
As shown in Eqs.~(\ref{so10}) and (\ref{24}), they should appear 
in the diagonal part of the matrix (\ref{so10/z2z2}).
The eigenvalues of $P$ and $P'$ are the imposed parities 
(or boundary conditions) of fields in the adjoint representations.   
The wave function of a field with parity $(+-)$, for instance,  
must vanish on the brane at $y=y_c/2$, and only fields assigned $(++)$ 
parities contain massless modes in their KK spectrum.   
%
%

An $N=2$ gauge multiplet consists of an $N=1$ vector multiplet  
($V^a=(A^{a}_\mu,\lambda^{1a})$) and an $N=1$ chiral
multiplet ($\Sigma^a=((\Phi^a+iA_5^a)/\sqrt{2},\lambda^{2a})$). 
In order to break $N=2$ SUSY to $N=1$, opposite parities
must be assigned to the vector and the chiral multiplets
in the same representation.  
The parities of $N=1$ vector multiplets coincide with the parities of
the corresponding $SO(10)$ generators in Eq.~(\ref{so10/z2z2}).
From the assigned eigenvalues in Eq.~(\ref{so10/z2z2}),
only the gauge multiplets associated with 
${\bf (8,1)}_0^{++}$, ${\bf (1,3)}_0^{++}$, and two ${\bf (1,1)}_0^{++}$,
which correspond to the $SU(3)_c\times SU(2)_L\times U(1)_Y\times U(1)_X$
generators, contain massless modes.
Therefore, at low energy the theory is effectively described by  
a $4D$ $N=1$ supersymmetric theory 
with $SU(3)_c\times SU(2)_L\times U(1)_Y\times U(1)_X$. 

As seen in Eq.~(\ref{so10/z2z2}), all of the gauge multiplets associated with  
the diagonal components ${\bf 24}_0$, ${\bf 1}_0$
in Eq.~(\ref{so10}) survive at B1.
Thus, on B1 $SU(5)\times U(1)_X$ is preserved \cite{hebecker}.
On the other hand, in Eq.~(\ref{so10/z2z2}), 
the elements with $(++)$ and $(-+)$ parities can compose a second (distinct) 
set of $SU(5)'\times U(1)_X'$ gauge multiplets.   
In the $SU(5)$ generator at B1,  
${\bf 24}_0$ ($={\bf (8,1)}_0^{++}+{\bf (1,3)}_0^{++}+{\bf (1,1)}_0^{++}
+{\bf (3,\overline{2})}_{-5/6}^{+-}+{\bf (\overline{3},2)}_{5/6}^{+-}$), the 
${\bf (3,\overline{2})}_{-5/6}^{+-}$ and ${\bf (\overline{3},2)}_{5/6}^{+-}$
are replaced by ${\bf (3,2)}_{1/6}^{-+}$ and
${\bf (\overline{3},\overline{2}})_{-1/6}^{-+}$ at B2
which belong in ${\bf 10}_{-4}$ and ${\bf \overline{10}}_{4}$ 
respectively of $SU(5)\times U(1)_X$,    
\begin{eqnarray}
{\bf 24}_0'={\bf (8,1)}_0^{++}+{\bf (1,3)}_0^{++}+{\bf (1,1)}_0^{++}
+{\bf (3,2)}_{1/6}^{-+}+{\bf (\overline{3},\overline{2}})_{-1/6}^{-+}~~~
{\rm at~B2}~, 
\end{eqnarray}
where the assigned hypercharges coincide 
with those given in ``flipped'' $SU(5)'\times U(1)_X'$~\cite{flipped}.    
The $U(1)_X'$ generator at B2 is defined as
\begin{eqnarray}
{\rm diag}({\bf 1}_{3\times 3}, -{\bf 1}_{2\times 2},
-{\bf 1}_{3\times 3}, {\bf 1}_{2\times 2}) ~.  
\end{eqnarray}
%
Thus, the $U(1)_X'$ charges of the surviving elements at B2
turn out to be zero, while the other components are
assigned $-4$ or $4$.
The $U(1)_X'$ generator and the matrix elements with $(++)$, $(-+)$ parities
in Eq.~(\ref{so10/z2z2}) 
can be block-diagonalized to the form in Eq.~(\ref{so10}) 
through the unitary transformation 
\begin{eqnarray}
U=\left(\begin{array}{c|ccc}
I_{3\times 3} & 0 & 0 & 0 \\ \hline
0 & 0 & 0 & I_{2\times 2} \\
0 & 0 & I_{3\times 3} & 0 \\
0 & I_{2\times 2} & 0 & 0
\end{array}\right)_{10\times 10} ~.
\end{eqnarray}
We conclude that the gauge multiplets surviving at B2 
are associated with a second (``flipped'') $SU(5)'\times U(1)_X'$  
embedded in $SO(10)$~\cite{flipped}.  
%

With opposite parities assigned to the chiral multiplets, 
two vector-like pairs 
$\Sigma_{{\bf (\overline{3},1)}_{-2/3}^{++}}$,
$\Sigma_{{\bf (3,1)}_{2/3}^{++}}$ and $\Sigma_{{\bf (1,1)}_{1}^{++}}$,
$\Sigma_{{\bf (1,1)}_{-1}^{++}}$ contain massless modes.
Although the non-vanishing chiral multiplets at B1 are
$\Sigma_{{\bf 10}_{-4}}$
($=\Sigma_{{\bf (\overline{3},1)}_{-2/3}^{++}}
+\Sigma_{{\bf (3,2)}_{1/6}^{+-}}
+\Sigma_{{\bf (1,1)}_{1}^{++}}$) and
$\Sigma_{{\bf \overline{10}}_{4}}$
($=\Sigma_{{\bf (3,1)}_{2/3}^{++}}
+\Sigma_{{\bf (\overline{3},\overline{2})}_{-1/6}^{+-}}
+\Sigma_{{\bf (1,1)}_{-1}^{++}}$),
$\Sigma_{{\bf (3,2)}_{1/6}^{+-}}$ and
$\Sigma_{{\bf (\overline{3},\overline{2})}_{-1/6}^{+-}}$ are replaced by
$\Sigma_{{\bf (3,\overline{2})}_{-5/6}^{-+}}$ and
$\Sigma_{{\bf (\overline{3},2)}_{5/6}^{-+}}$
at B2 that are contained in $\Sigma_{{\bf 24}_0}$ and 
$\Sigma_{{\bf \overline{24}}_0}$ at B1.
Together with the vector-like pairs containing massless modes,
they compose ${\bf 10}_{-4}'$ and ${\bf \overline{10}_4}'$-plets  
of $SU(5)'\times U(1)_X'$, 
\begin{eqnarray}
&&\Sigma_{{\bf 10}_{-4}'}
=\Sigma_{{\bf (\overline{3},1)}_{-2/3}^{++}}+
\Sigma_{{\bf (3,\overline{2})}_{-5/6}^{-+}}
+\Sigma_{{\bf (1,1)}_{-1}^{++}}~,~~  \\
&&\Sigma_{{\bf \overline{10}}_{4}'}
=\Sigma_{{\bf (3,1)}_{2/3}^{++}}+\Sigma_{{\bf (\overline{3},2)}_{5/6}^{-+}}
+\Sigma_{{\bf (1,1)}_{1}^{++}} ~~~~~~~~~{\rm at ~ B2} ~.
\end{eqnarray}

Now let us discuss the $N=2$ (bulk) hypermultiplet $H$ $(=(\phi,\psi))$, 
$H^c$ ($=(\phi^c,\psi^c)$) in the vector representations ${\bf 10}$, 
${\bf 10^c}$ ($={\bf 10}$) of $SO(10)$, where   
$H$ and $H^c$ are $N=1$ chiral multiplets.  
Under $SU(5)\times U(1)_X$ and $SU(3)_c\times SU(2)_L\times U(1)_Y$,
${\bf 10}$ and ${\bf 10^c}$ are 
\begin{eqnarray}
{\bf 10}=
\left(\begin{array}{c}
{\bf 5}_{-2} \\
\hline
{\bf \overline{5}}_{2}
\end{array}\right)=
\left(\begin{array}{c}
~{\bf (3,1)}_{-1/3}^{+-}  \\
{\bf (1,2)}_{1/2}^{++}  \\
\hline
{\bf (\overline{3},1)}_{1/3}^{-+}  \\
~{\bf (1,\overline{2})}_{-1/2}^{--}
\end{array}\right)~, ~~
{\bf 10^c}=
\left(\begin{array}{c}
{\bf 5^c}_{2} \\
\hline
{\bf \overline{5^c}}_{-2}
\end{array}\right)=
\left(\begin{array}{c}
{\bf (3^c,1)}_{1/3}^{-+}  \\
~{\bf (1,2^c)}_{-1/2}^{--}  \\
\hline
~{\bf (\overline{3}^c,1)}_{-1/3}^{+-}  \\
{\bf (1,\overline{2}^c)}_{1/2}^{++}
\end{array}\right) ~,
\end{eqnarray}
where the subscripts $\pm 2$ are $U(1)_X$ charges and the remaining subscripts 
indicate the hypercharges $Y$.
The superscripts on the matrix elements denote the eigenvalues of
the $P$ and $P'$ operations.
As in the $N=2$ vector multiplet, 
opposite parities must be assigned for $H$ and $H^c$
to break $N=2$ SUSY to $N=1$.   The massless modes are contained 
in the two doublets ${\bf (1,2)}_{1/2}^{++}$ and
${\bf (1,\overline{2^c})}_{1/2}^{++}$ ($={\bf (1,2)}_{1/2}^{++}$).
While the surviving representations at B1,
${\bf (3,1)}_{-1/3}^{+-}$ and ${\bf (1,2)}_{1/2}^{++}$
(also ${\bf (\overline{3}^c,1)}_{-1/3}^{+-}$ and
${\bf (1,\overline{2}^c)}_{1/2}^{++}$) compose two ${\bf 5}_{-2}$
(or ${\bf \overline{5}^c}_{-2}$) of $SU(5)\times U(1)_X$,
at B2  the non-vanishing representations are
two ${\bf \overline{5}}_{2}'$ of $SU(5)'\times U(1)_X'$,
\begin{eqnarray}
{\bf \overline{5}}_{2}' ={\bf (\overline{3},1)}_{1/3}^{-+}
+{\bf (1,2)}_{1/2}^{++} ~~\bigg({\rm or}~~{\bf (3^c,1)}_{1/3}^{-+}
+{\bf (1,\overline{2}^c)}_{1/2}^{++}\bigg) ~~{\rm at~B2}~.
\end{eqnarray}

In this model, the $SU(2)$ $R$-symmetry which generally exists in $N=2$ 
supersymmetric theories is explicitly broken to $U(1)_R$.  
Since $N=1$ SUSY is present on both branes,
$U(1)_R$ symmetry should be respected. 
We note that different $U(1)_R$ charges can be assigned
to $H_{{\bf 10}_{-4}}$ and $H^c_{{\bf 10^c}_{4}}$  
as shown in Table I \cite{nomura}.    
%
%
\vskip 0.6cm
\begin{center}
\begin{tabular}{|c||ccc|c|} \hline
$U(1)_R$&  & $V$,~ $\Sigma$ & & $H$, $H^c$
\\
\hline \hline
$1$ & & & & $\phi^c$
\\
$1/2$ & & $\lambda^1$& & $\psi^c$
\\
$0$ & ~~$A_\mu$& & $\Phi$, $A_5$~~ & $\phi$
\\
$-1/2$ & & $\lambda^2$& & $\psi$
\\
\hline 
\end{tabular}
\vskip 0.4cm
{\bf Table I.~}$U(1)_R$ charges of the vector and hypermultiplets.
\end{center}

To construct a realistic model, which includes inflation, based on 
$SU(3)_c\times SU(2)_L\times U(1)_Y\times U(1)_X$,  
we introduce a $U(1)_{PQ}$ axion symmetry and 
``matter'' parity $Z_2^m$~\cite{khalil}.  
For simplicity, let us assume that the MSSM matter superfields as well as 
the right-handed neutrinos are brane fields residing at B1.\footnote{If 
the first two quark and lepton families reside on B2 where
$SU(5)'\times U(1)_X'$ is preserved, undesirable mass relations between   
the down-type quarks and the charged leptons do not arise.
Mixings between the first two and the third families
can be generated by introducing bulk superheavy hypermultiplets 
in the spinor representations of $SO(10)$~\cite{KS2}.  }  
They belong in ${\bf 10}_i$, ${\bf \overline{5}}_i$,
and ${\bf 1}_i$ of $SU(5)$, where $i$ is the family index.
Their assigned $U(1)_X$, $U(1)_R$ and $U(1)_{PQ}$ charges and matter parities 
appear in Table II.  
\vskip 0.6cm
\begin{center}
\begin{tabular}{|c||c|c|c|c|c|c|c|c|} \hline
Fields & $S$& $~N_H~$& $~\overline{N}_H~$ &$~{\bf 10}_B^{(')}~$&
$~{\bf \overline{10}}_B^{(')}~$ & ${\bf 1}_i$  &
$~~{\bf \overline{5}}_i~~$ & $~{\bf 10}_i~$
\\ \hline \hline
$X^{(')}$ & $0$& $5$& $-5$& $-4$& $4$& $5$& $-3$& $1$
\\
$R$ & $1$ & $0$& $0$& $0$ & $0$ & $1/2$ & $1/2$ & $1/2$
\\
$PQ$ & $0$& $0$& $0$& $0$ & $0$ & $0$ & $-1$ & $-1/2$
\\
$Z_2^m$ & $+$& $+$& $+$& $-$& $-$& $-$& $-$& $-$
\\
\hline \hline
Fields & ~$h_1^{(')}$~ & $h_2^{(')}$ & $~\overline{h}_1^{(')}~$ &
$~\overline{h}_2^{(')}~$ & $\Sigma_1$ & $\Sigma_2$ & $\overline{\Sigma}_1$ &
$\overline{\Sigma}_2$
\\
\hline \hline
$X^{(')}$& $-2$& $-2$& $2$ & $2$ & $0$ & $0$& $0$ & $0$
\\
$R$ & $0$& $1$& $0$ & $1$& $1/2$ & $1/2$ & $0$ & $0$
\\
$PQ$  & $1$& $1$& $3/2$ & $3/2$& $-1$ & $-3/2$ & $1$ & $3/2$
\\
$Z_2^m$& $+$& $-$& $+$& $-$& $+$& $+$& $+$& $+$
\\
\hline
\end{tabular}
\vskip 0.4cm
{\bf Table II.~}$U(1)_X^{(')}$, $U(1)_R$, $U(1)_{PQ}$ charges
and matter parities of the superfields.
\end{center}

We introduce two pairs of hypermultiplets 
$(H_{\bf 10},H^c_{\bf 10^c})$ and 
$(H_{\bf \overline{10}},H^c_{\bf \overline{10}^c})$ 
($=(H_{\bf 10},H^c_{\bf 10^c})$) in the bulk.
The two $SU(5)$ Higgs multiplets $h_1$ and $\overline{h}_1$ (${\bf 5}$ and
${\bf \overline{5}}$) arise from $H_{\bf 10}$ and $H_{\bf \overline{10}}$,  
and their $U(1)_R$ charges are chosen to be zero. 
As discussed above, 
the $N=2$ superpartners $H_{\bf 10}^c$ and $H_{\bf \overline{10}}^c$
also provide superfields $h_2$ and $\overline{h}_2$
with ${\bf 5}^{(')}$ and ${\bf \overline{5}}^{(')}$ representations at B1 (B2). 
However, their $U(1)_R$ charges are unity unlike $h_1$ and $\overline{h}_1$.
To make them superheavy
%
we can introduce another pair of ${\bf 5}$ and ${\bf \overline{5}}$
with zero $U(1)_R$ charges and `$-$' matter parities on the brane.
%
%
%

The superpotential at B1, neglecting the superheavy particles' contributions 
except for the inflatons, is given by
\begin{eqnarray} \label{3211}
W&=&\kappa S\bigg(N_H\overline{N}_H
%
-M^2\bigg)
+\frac{\sigma_1}{M_P}\Sigma_1\Sigma_2h_1\overline{h}_1
+\frac{\sigma_2}{M_P}\Sigma_1\Sigma_2\overline{\Sigma}_1\overline{\Sigma}_2
 \\
&&+y^{(d)}_{ij}{\bf 10}_i{\bf 10}_jh_1+
y^{(ul)}_{ij}{\bf 10}_i{\bf \overline{5}}_j\overline{h}_1
+y^{(n)}_{ij}{\bf 1}_i{\bf \overline{5}}_jh_1
+\frac{y^{(m)}_{ij}}{M_P}{\bf 1}_i{\bf 1}_j\overline{N}_H\overline{N}_H ~,
 \nonumber
\end{eqnarray}
where $S$, $N_H$, $\overline{N}_H$, $\Sigma_{1,2}$, and 
$\overline{\Sigma}_{1,2}$ are singlet fields. 
Their assigned quantum numbers appear in Table II.
While $S$, $N_H$, $\overline{N}_H$, and $h_1$, $\overline{h}_1$ are 
bulk fields, the rest are brane fields residing on B1.  
$N_H$, $\overline{N}_H$ should be embedded
in ${\bf 16}_H$, ${\bf \overline{16}}_H$, and 
the other components in ${\bf 16}_H$, ${\bf \overline{16}}_H$ could 
be made heavy by pairing them with proper brane fields.   
From Eq.~(\ref{3211}),
it is straightforward to show that the
SUSY vacuum corresponds to $\langle S\rangle=0$, while
$N_H$ and $\overline{N}_H$ develop VEVs of order $M$.   
(After SUSY breaking in the manner of $N=1$ SUGRA, 
$\langle S\rangle$ acquires a VEV
of order $m_{3/2}$ (gravitino mass)).  
They break $SU(3)_c\times SU(2)_L\times U(1)_Y\times U(1)_X$ 
to the MSSM gauge group, and make the massless modes 
in $\Sigma_{{\bf 10}_{-4}}$ ($\equiv {\bf 10}_{B}$) and 
$\Sigma_{{\bf \overline{10}}_{4}}$ ($\equiv {\bf \overline{10}}_B$)
superheavy~\cite{KS2}.  
From the last term in Eq.~(\ref{3211}), the VEV of $\overline{N}_H$ 
also provides masses to the right-handed Majorana neutrinos.     
%
%

Because of the presence of the soft terms,
$\Sigma_{1,2}$ and $\overline{\Sigma}_{1,2}$, which carry $U(1)_{PQ}$ charges, 
can obtain intermediate scale VEVs of order $\sqrt{m_{3/2}M_P}$.
They lead to a $\mu$ term of order $m_{3/2}$ in MSSM  
as desired \cite{mu, mu2}.
Of course, the presence of $U(1)_{PQ}$ also resolves 
the strong CP problem~\cite{axion}.    
%
%

The Higgs fields $h_1$ and $\overline{h}_1$ contain color triplets 
as well as weak doublets.
Since the triplets in $h_1$ and $\overline{h}_1$ are just superheavy KK modes,
a small coefficient ($\mu\sim$ TeV) accompanying $h_1\overline{h}_1$  
more than adequately suppresses dimension 5 operators that induce proton decay.
Proton decay can still proceed via superheavy gauge bosons with masses 
$\approx \pi/y_c$ and are adequately suppressed ($\tau_p\sim 10^{34-36}$ yr). 
 

\section{Inflation and Leptogenesis}

The first two terms in the superpotential (\ref{3211}) are ideally suited 
for realizing an inflationary scenario along the lines 
described in Refs.~\cite{KS,hybrid,khalil}.  
We will not provide any details except to note that the breaking of $U(1)_X$ 
takes place near the end of inflation which can lead to the appearance of 
cosmic strings. Since the symmetry breaking scale of $U(1)_X$ is 
determined from inflation to be close to $10^{16}$ GeV \cite{hybrid}, 
the cosmic strings are superheavy and therefore not so desirable 
(because of potential conflict with the recent $\delta T/T$ measurements.) 
They can be simply avoided by following the strategy outlined 
in Ref.~\cite{khalil}, in which suitable non-renormalizable terms are added 
to Eq.~(\ref{3211}), 
such that $U(1)_X$ is broken along the inflationary trajectory and the strings 
are inflated away.  Remarkably, the salient features of the inflationary 
scenario are not affected by the addition of such higher order terms.  
%
%
For $\kappa$ somewhat smaller (larger) than 
$10^{-3}$, $n$ varies between 0.98 and 0.99.   

The inflationary epoch ends with the decay of the oscillating fields 
$S$, $N_H$, $\overline{N}_H$ into the MSSM singlet (right-handed) neutrinos, 
whose subsequent out of equilibrium decays yield the observed 
baryon asymmetry via leptogenesis~\cite{lepto,ls}.  
The production of the right-handed neutrinos and sneutrinos proceeds via
the superpotential couplings $\nu^c_i\nu^c_j\overline{N}_H\overline{N}_H$ 
and $SN_H\overline{N}_H$ on B1, where $\nu^c_{i,j}$ ($i, j=1,2,3$)  
denote the $SU(5)$ singlet (right-handed neutrinos) 
carrying non-zero $U(1)_X$ charges.  
Taking account of the atmospheric and solar neutrino oscillation 
data~\cite{nuoscil}, 
and assuming a hierarchical pattern of neutrino masses 
(both left- and right-handed ones), it turns out that the inflaton decays 
into the lightest (the first family right handed neutrino)~\cite{KS,pati}.  
The discussion proceeds along the lines discussed in Ref.~\cite{lasymm}, 
where it was shown that a baryon asymmetry of the
desired magnitude is readily obtained.

The 5D inflationary solution requires positive vacuum energies on both branes
B1 and B2~\cite{KS}.  While inflation could be driven by the first two terms 
in Eq.~(\ref{3211}) at B1,
an appropriate scalar potential on B2 is also necessary.
Since the boundary conditions in Eqs.~(\ref{bdy1}) and (\ref{bdy2})
require $\Lambda_1$ and $\Lambda_2$ to simultaneously vanish,
it is natural for $S$ to be a bulk field.
The VEVs of $S$ on the two branes can be adjusted
such that the boundary conditions are satisfied.
As a simple example, consider the following superpotential at B2,
\begin{eqnarray} \label{b1superpot}
W_{B2}=\kappa_2S(Z\overline{Z}-M_2^2) ~,
\end{eqnarray}
where $Z$ and $\overline{Z}$ are $SO(10)$ singlet superfields 
with opposite $U(1)_R$ charges.
Thus, only the gauge symmetry $U(1)_X$ on B1 is broken during inflation.  

\section{Conclusion}

Inspired by recent attempts to construct realistic 5D SUSY GUT models,
we have presented a realistic inflationary scenario in this setting, 
along the lines proposed in Ref.~\cite{KS}.  
Inflation is implemented through $F$-term scalar potentials on the two branes, 
which is allowed by 5D Einstein gravity.  
We have discussed the transition from the inflationary 
to the radiation dominated phase, and
provided a realistic 5D SUSY $SO(10)$ model in which the compactification 
on an $S^1/(Z_2\times Z_2')$ orbifold leads to the gauge symmetry 
$SU(3)_c\times SU(2)_L\times U(1)_Y\times U(1)_X$.   
Inflation is associated with the breaking of the gauge symmetry $U(1)_X$, 
at a scale very close to $M_{\rm GUT}$.   
Baryogenesis via leptogenesis is very natural in this approach, 
and the scalar spectral index $n=0.98-0.99$.   
The gravitational contribution to the quadrupole
anisotropy is found to be tiny ($\lapproxeq 10^{-4}$).  

\vskip 0.3cm
\noindent {\bf Acknowledgments}

\noindent
We acknowledge helpful discussions with S. M. Barr and Tianjun Li.    
The work is partially supported 
by DOE under contract number DE-FG02-91ER40626.

\end{document}